# Epitaxial growth of bilayer Bi(110) on two-dimensional ferromagnetic $Fe_3GeTe_2$

*Yilian Xi[1,2][†], Mengting Zhao[1,2][†], Haifeng Feng[1][*], Ying Sun[1], Xingkun Man[1], Xun Xu[2][*], Weichang Hao[1], Shi Xue Dou[2], Yi Du[1][*]*

[1]School of Physics, Beihang University, Beijing 100191, China
[2]Institute for Superconducting and Electronic Materials (ISEM) and BUAA-UOW Joint Research Centre, Australian Institute for Innovative Materials (AIIM), University of Wollongong, Wollongong, NSW 2500, Australia

**Abstract**

Heterostructures of two-dimensional (2D) layered materials with selective compositions play an important role in creating novel functionalities. Effective interface coupling between 2D ferromagnet and electronic materials would enable the generation of exotic physical phenomena caused by intrinsic symmetry breaking and proximity effect at interfaces. Here, epitaxial growth of bilayer Bi(110) on 2D ferromagnetic $Fe_3GeTe_2$ (FGT) with large magnetic anisotropy has been reported. Bilayer Bi(110) islands are found to extend along fixed lattice directions of FGT. The six preferred orientations could be divided into two groups of three-fold symmetry axes with the difference approximately to 26°. Moreover, *dI/dV* measurements confirm the existence of interface coupling between bilayer Bi(110) and FGT. A variation of the energy gap at the edges of bilayer Bi(110) is also observed which is modulated by the interface coupling strengths associated with its buckled atomic structure. This system provides a good platform for further study of the exotic electronic properties of epitaxial Bi(110) on 2D ferromagnetic substrate and promotes potential applications in the field of spin devices.

## 1. Introduction

Bismuth, as the last element of the nitrogen group, has become a research hotspot in recent decades, owing to its dramatically exotic physical properties and chemical properties, and its widely developed applications in many fields [1-3]. For example, bismuth is a promising anode material for sodium-ion batteries due to its high capacity and suitable operating potential. The electrochemical reduction of carbon dioxide has the potential to directly convert renewable electricity into valuable chemical raw materials or fuels, thereby closing the anthropogenic carbon cycle, in which bismuth as a highly efficient electrocatalyst for carbon dioxide reduction reaction plays an important role in the field of catalysis [4-10]. Bismuth can also be used as a strain-independent Infrared reducer optoelectronic material for biophotonics [11]. Semimetal bismuth is imperative to condensed matter physics due to its exotic intrinsic properties such as high atomic number, long Fermi wavelength as well as strong intrinsic spin-orbital coupling (SOC) [12-16], which endow bismuth great potential in the field of the electronic and spintronic devices.

Inspired by the discovery of graphene [17], the investigations into two-dimensional (2D) materials have undergone a burgeoning growth [18-22]. The electronic properties of ultrathin films are completely different from those of their bulk counterparts owing to size effects and/or structure changes. These effects are particularly pronounced for Bi thin films, because their special electronic properties are balanced between metals and semiconductors. Stacking different layered 2D crystals into heterostructure opens an opportunity to create new materials and control their electronic properties, offering unprecedented freedom in combining different materials as well as providing novel physics. In recent years, ultra-thin Bi films have been successfully grown on various metallic and semiconducting substrates, in the form of either Bi(110) or Bi(111) depended on the growth conditions and coverages [23-27]. Bilayer Bi(111) films, also known as bismuthene, have become a research hotspot in recent years as a high-temperature quantum spin Hall material [28-30]. Metal to semiconductor transition on ultra-thin Bi(110) film has been reported due to edge reconstructions, in which a large energy gap of 0.4 eV makes it appeal in practical applications as electronic devices in nano-scale [16]. Topological properties have also been identified on Bi(110) films grown on highly oriented pyrolytic graphite (HOPG) substrate, which exhibit sensitivity towards the interaction between the film and its substrate [31]. Meanwhile, it has been recognized that the interface coupling between the Bi(110) film and its substrate could influence the electronic properties of the system through determining the atomic structure, interface charge transfer, orbital hybridization and proximity effect, as exemplified by the unusual superconducting (SC) proximity effect on Bi(110) grown on NbSe$_2$ substrate [32]. Thus, growing ultra-thin Bi films on functional substrates and engineering the interface coupling would benefit for searching exotic physical phenomena and potential applications.

The recent discovery of intrinsic 2D ferromagnetism in layered materials opens up unprecedented opportunities for exotic phenomena and device applications [33-38]. Among these, 2D ferromagnetic materials, Fe$_3$GeTe$_2$ (FGT) is significantly promising due to high Curie temperature ($T_c$) in bulk, which could be tuned to room temperature by using ion gating in the 2D form [34]. Considering the itinerant ferromagnetic property of FGT, the large magnetic anisotropy and strong SOC suppress thermal fluctuations, leading to long-range magnetic order. Moreover, as magnetism is based on strong short-range correlations between electron spin and orbital degrees of freedom that are inherently changed at the interfaces [39], the interface coupling between 2D ferromagnetic materials and epitaxially grown thin films could not only modify its magnetic properties but also the physical properties of adjacent nonmagnetic layers, especially for those materials with strong SOC including Bi ultra-thin films.

Based on the above-mentioned motivations, in this work we have grown sub-monolayer bilayer Bi(110) islands on FGT substrate which is a robust 2D ferromagnet with strong magnetic anisotropy. Through scanning tunnelling microscopy and spectroscopy (STM and STS) studies, it is found that the islands with different diameters appear randomly over the surface. By comparing these bilayer Bi(110) islands atomic lattices, it is determined that the islands extend along the shorter side of the unit cell of Bi(110). Furthermore, the bilayer Bi(110) islands extend along six preferred orientations approximately ±13° from the [100] direction of the FGT(001) growth substrate. Importantly, the *dI/dV* spectra reveals that effective interface couplings exist in both the electronic band structure and magnetic properties of the systems. This work provides insights of the effective interface coupling in the heterostructure of 2D ferromagnet and Bi ultra-thin films at atomic scale.

## 2. Experimental detail

High-quality single crystals of FGT were grown by using chemical transport method with iodine as the transport agent according to Ref. 40. High-purity Fe (99.99%), Ge (99.99%), and Te (99.99%), in a stoichiometric ratio of 3:1:2 with iodine as the transport agent were sealed in an evacuated quartz tube. Powder X-ray diffraction (XRD, PANalytical X9 PertPro X-ray diffractometer using Cu Kα radiation) was performed to investigate the structure of FGT samples. Room-temperature Raman measurements were performed on an In Via Reflex Raman spectrometer with a laser at 532 nm. The surface morphologies of the GeH samples were investigated using scanning electron microscopy (SEM, JEOL JSM-7500FA). Besides, the elemental analysis was performed on an energy dispersive X-ray spectroscopy (EDS, OXFORD INCA X-ACT).

STM and STS measurements were performed using a low-temperature STM system (STM1500, Unisoku Co.) under ultrahigh vacuum (UHV) at 4.2 K. An FGT single crystal was cleaved with scotch tape to obtain a shiny mirror-like surface under vacuum of better than $5.0 \times 10^{-10}$ Torr. And then, high-purity Bi (99.99%) was deposited onto the surface of FGT at room temperature using a standard Knudsen cell. All measurements were performed at 4.2 K. All the STM images were obtained in constant current mode. The STS differential conductance (*dI/dV*) was obtained with lock-in detection by applying modulation to the tunnel at 613 Hz. All the STM images were analyzed by WSxM software [41].

## 3. Results and Discussion

FGT is a Van der Waals (vdW) metallic ferromagnet with the space group *P63/mmc* [42, 43]. Figure 1a and Figure 1b show the top-view and side-view crystal structures of FGT, respectively. FGT monolayer consists of covalently bonded $Fe_3Ge$ layer and Te layers above and below the $Fe_3Ge$ layer. In an $FeGe_3$ layer, there are two inequivalent Wyckoff Fe sites denoted as $Fe_1$ and $Fe_2$. Due to the weak vdW bonding, FGT flakes easily exfoliated from the bulk crystals have the plane surfaces parallel to the $Fe_3Ge$ layers. Figure 1c shows the X-ray pattern of grown FGT crystal, in which all peaks can be indexed to the space group *P63/mmc*. It clear that the peaks only show the *c* axis normal with exclusively 00l diffraction, indicating that the grown FGT is a high-quality single crystal. The as-grown single crystals are prepared with lateral dimensions up to several millimetres by using chemical transport method (inset Figure 1c). The Raman spectrum of FGT sample in Figure 1d shows three phonon peaks at 122.21 cm$^{-1}$ ($E_{2g}^1$ vibrational mode), 143.27 cm$^{-1}$ ($A_{1g}^1$ vibrational mode) and 268.85 cm$^{-1}$ ($A_{1g}^2$ vibrational mode) at room temperature. These three Raman modes are illustrated in the insets of Figure 1d. The Raman mode of $E_{2g}^1$ represents the in-plane vibration including all atoms, while the Raman modes of $A_{1g}^1$ and $A_{1g}^2$ respectively show Te and Fe atoms vibration along the out-of-plane direction [44]. Figure 1e illustrates the scanning electron microscopy (SEM) morphology of FGT surface. Elemental mapping is also carried out by energy dispersive X-ray spectroscopy (EDS) to investigate the elemental dispersion on the FGT surface. Figures 1f-h respectively show the elements Fe, Ge and Te distributing uniformly on the FGT surface. All the above characterizations confirm the high quality and purity of the FGT single crystal.

In our experiment, FGT as the substrate is used to support a layer-by-layer growth of Bi(110) islands. The schematic illustration of our experimental setup is illustrated in Figure S1a, in which Bi atoms are evaporated from a standard Knudsen cell onto an FGT surface at room temperature. The coverage of Bi on the surface of FTG is controlled by the depositing time at a constant Bi flux. The samples are then cooled to 4.2 K for STM and STS measurements (detailed in the Experimental Section). Due to Bi is a heavy metal atom with strong SOC, remarkable interplay between spin-spin and spin-orbital at the interface with FGT substrate is expected to happen as illustrated in Figure S1b, which is highly desired in developing new spintronic devices.

The evolution of the topographies of sub-monolayer Bi on FGT substrate are characterized by low-temperature STM system. As shown in Figure 2a, at the initial stage of growth, the Bi clusters randomly distribute on the FGT substrate (STM images of bare FGT substrate are shown in Figure S2). The corresponding height profile of several Bi clusters is shown in Figure 2d. With the coverage increasing, flat Bi islands appear on the surface as shown in Figure 2b, whose size further increases in the in-plane directions to several tens of nanometers at even higher Bi coverage as shown in Figure 2c. Those isolated Bi islands with different sizes show an identical thickness. As shown in the height profiles in Figure 2e and Figure 2f, the apparent heights are around 0.69 nm which is consistence with bilayer Bi(110) [16, 45]. In fact, according to pervious research the freestanding ultra-thin Bi(110) islands prefer to form a bilayer structure due to their dangling bonds. As the outermost shell electron configuration of Bi is $6s^26p^3$, each atom in the bulk Bi forms three σ bonds with its nearest neighbors. At the low coverage, a ladder structure with dangling bonds is formed, thus preferring to adopt the bilayer (or even monolayer) stacking mode in freestanding ultra-thin Bi(110) islands [31,46].

Figure 3a shows the representative STM image of the bilayer Bi islands on FGT substrate in the sub-monolayer regime. The atomically resolved STM image of Bi(110) thin film is shown in Figure 3b, in which a rectangle with crystal lattice constants of 4.73 Å × 4.52 Å along the long edge as indicated by the arrow line. Meanwhile, it is found that almost all the angles between these sharp long edges and short edges are 145°. Thus, the edges are determined to terminate along the unit cell direction (*b* direction for the long edges) or the diagonal direction of the unit cell. The top-view and side-view crystal structures of Bi(110) are illustrated in Figure 3d. Each bilayer Bi(110) contains two bonded Bi atom sublayers forming a layered structures normal to its plane. The surface atoms form a rectangular unit cell with the crystal lattice constants of 4.75 Å × 4.54 Å, consistent with previously mentioned STM atomic structure. Moreover, the islands extend along the shorter side of the unit cell of Bi(110), similar phenomena has been observed in Bi(110) islands grown on Si(111) √3×√3-B substrate [47]. Figure 3c illustrates the detailed atomic network of FGT surface, clearly showing that the top Te-terminal surface has a hexagonal lattice structure (4.00 Å). By comparing these Bi(110) atomic lattices with the structure of a neighboring FGT(001) growth substrate, the angle between the shorter length of Bi(110) lattice and the exposed Te-terminal surface in FGT is 73° (Figure 3c ). Figure 3e exhibits one of stacking configurations between a bilayer Bi(110) and FGT(001) substrate. By comparing tremendous Bi(110) structures with the atomic lattice of the neighboring FGT(001) growth substrate, it is found that the Bi(110) islands extend along six preferred orientations, as shown in Figure 3f. It can be seen that these six preferred orientations could be divided into two groups of three-fold symmetry axes with the difference approximately to 26°. The similar phenomena also reported at Bi(110) on Si(111) √3×√3-B substrate [47,48]. The fixed directions of epitaxially grown bilayer Bi(110) islands imply the existence of necessary interface coupling between bilayer Bi(110) islands and FGT substrate.

Effective interface coupling on the electronic structures between Bi(110) and FGT is also confirmed in the STM/STS characterizations. Figure 4a shows the typical STM image of a bilayer Bi(110) island. It is found that the island has a ribbon-shape with a flat terrace in the middle and hump-like reconstructions along the long edges. Figure 4b shows the enlarged STM image at the edge to emphasize the subtle reconstructions. A line profile of hump-like superstructures at the edge of Bi (110) island (blue line in Figure 4b) is illustrated in Figure 4c, which is a 4 × 1 reconstruction. This phenomena is similar to the Bi(110) island grown on HOPG substrate caused by strain energy relaxation [16]. Figure 4d exhibits typical *dI/dV* curves measured at different positions, as indicated in Figure 4a. The density of states near the Fermi level of FGT is higher than that of bilayer Bi(110) island and edges, which is consistent to the semimetal properties of bismuth. In addition, several features of Bi island and FGT substrate share similarities in their *dI/dV* curves, indicating effective interface coupling in their electronic structures. For example, a prominent peak at a sample bias −60 mV, that originates from Fe *3d* orbitals [38], is identified on both Bi island and FGT. Moreover, for the two edges further suppression of the density of states near the Fermi level is identified, which agrees with the gap opening behavior due to structure reconstruction. Interestingly, it is found that the gaps at the two edges show an apparent shift as can been seen more clearly in the line map shown in Figure 4e (along the white arrow line in Figure 4a). As the position of the gap does not vary along the long edges (Figure S3 and Figure S4), this shift of the gap is very likely caused by the different interface coupling strength with FGT at the two edges, which origins from the buckled atomic structure (Figure 3d). A similar effect of the buckled atomic structure on the interface coupling has been observed on bilayer Bi(111). The variation of the interface coupling caused by buckled atomic structure of bilayer Bi(110), thus, not only demonstrates its existence but also suggests the feasibility of engineering the electronic structure of this system by adjusting its atomic structure.

## 4. Conclusion

In summary, atomically flat bilayer Bi(110) islands have been successfully grown on 2D ferromagnetic FGT substrate. The structural and electronic properties of this system are investigated by low-temperature STM-STS experiments. In the STM and STS studies, the effective interface coupling between bilayer Bi(110) and FGT is demonstrated through characterizing their atomic structures and local electronic structures. We observe that the islands have six preferred orientations approximately ±13∘ from the [100] direction of the FGT(001) growth substrate. In addition, a 4 × 1 reconstruction is observed on the long edges of bilayer Bi(110), whose energy gap is sensitive to the atomic buckled structure. The experimental realization of effective interface coupling between 2D ferromagnet and bismuth ultra-thin structures would promote the insights into the understanding of heterostructure in this system and the possible applications of spintronic devices.

**Acknowledgments**

The authors acknowledge the financial support from Beijing Natural Science Foundation (Z180007), the National Natural Science Foundation of China (12074021, 11874003, 12104033), the National Key Research & Development Program of China (2018YFE0202700), Chinese Postdoctoral Science Foundation (2021M690306) and Australian Research Council (ARC) (LP180100722, FT180100585). The authors acknowledge the supports from the BUAA-UOW Joint Research Centre.

**Figures and captions**

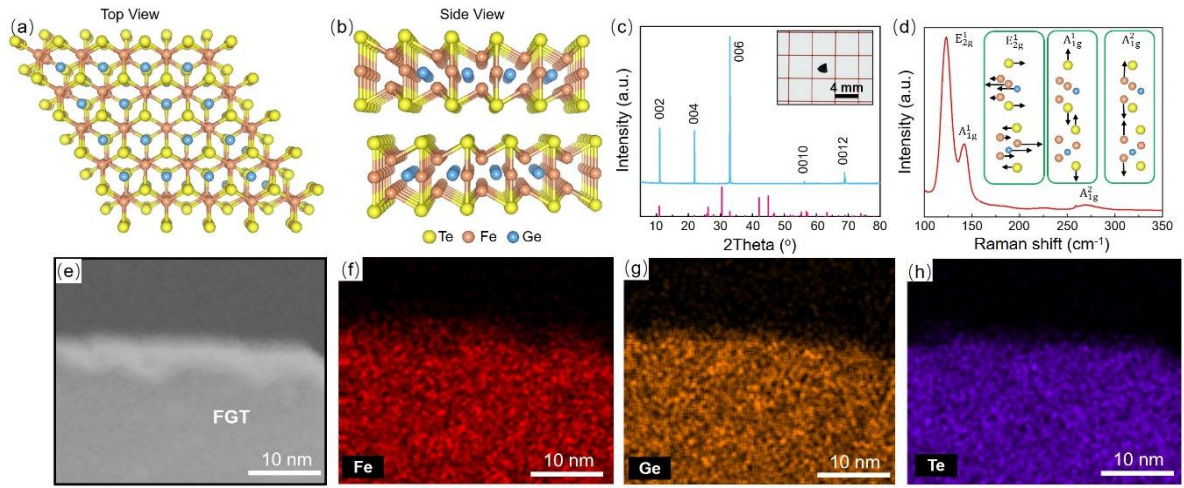

**Figure 1**. (a) The top-view crystal structure of FGT. (b) The side-view crystal structure of FGT. (c) X-ray diffraction pattern obtained from the cleaved plane of a FGT crystal at room temperature. The red line is the standard location and intensity, and the blue line is the measured data. The image of a representative single crystal is shown in the inset. (d) The Raman spectrum of grown FGT collected at room temperature. The insets represent the in-plane mode of $E_{2g}^1$ and the out-of-plane modes of $A_{1g}^1$ and $A_{1g}^2$. (e) The SEM image of FGT nanosheet on conductive substrate. (f)-(h) respectively show EDS elemental mapping images of Fe, Ge and Te atoms, corresponding to rectangle in (e).

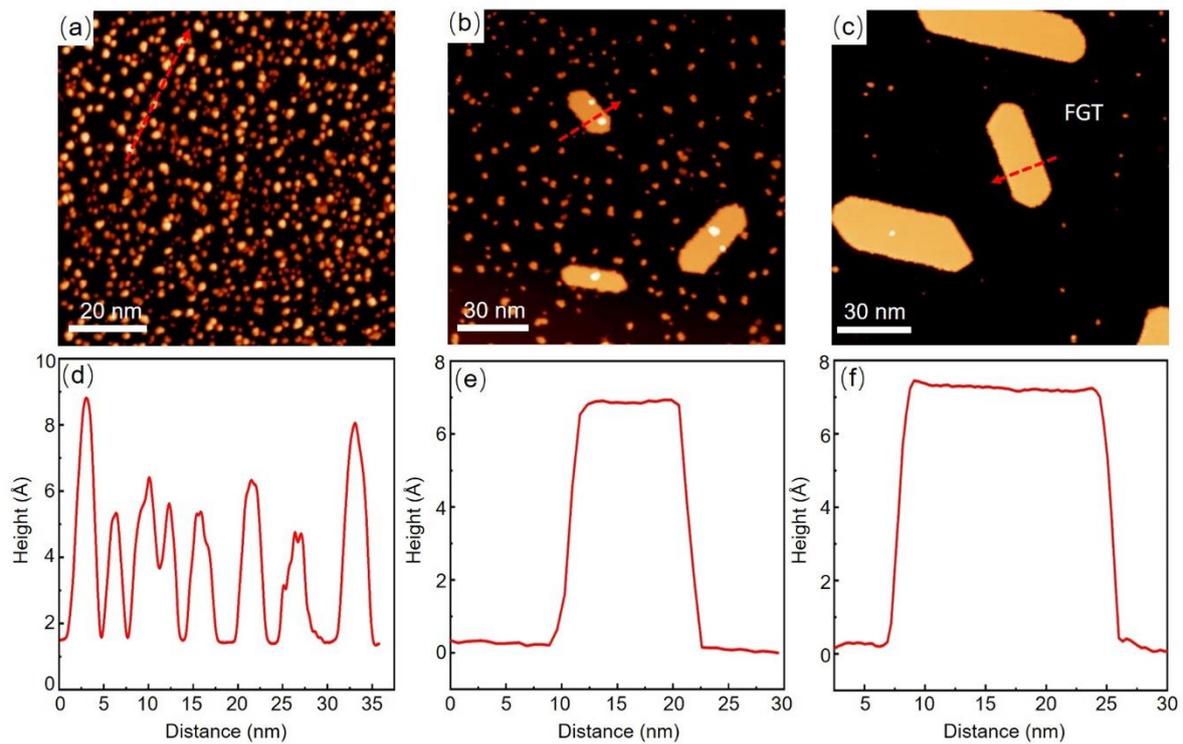

**Figure 2**. STM images of different Bi coverage on FGT surface. (a) Bi cluster, (b) growth Bi islands, (c) larger Bi islands. (d)–(f) show the corresponding line-scan profiles taken along the red dashed arrows in (a) to (c), respectively.

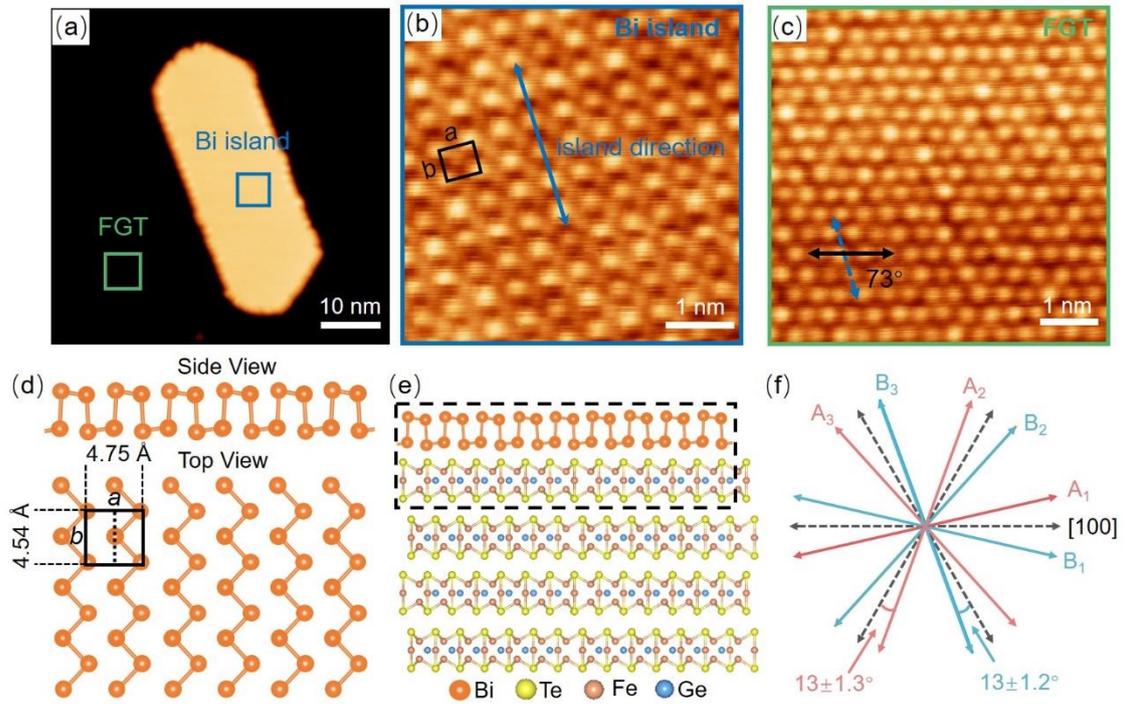

**Figure 3**. (a) Large scale STM image of the bilayer Bi(110) island on FGT surface. (b) The atomic resolved STM topography of a bilayer Bi(110) island terrace from the blue square in (a). (c) The atomic resolved STM topography of a neighboring FGT(001) substrate from the green square in (a), where the angle between short length of Bi crystal lattice (blue dashed arrow line) and the Te atom chains of FGT surface (black solid arrow line) is 73°. (d) Schematic illustration of side- and top-views of a bilayer Bi(110) crystal structure. (e) Side view of bilayer Bi(110) stacked on FGT substrate. (f) The schematic preferred orientation of bilayer Bi(110) islands grown on FGT substrate.

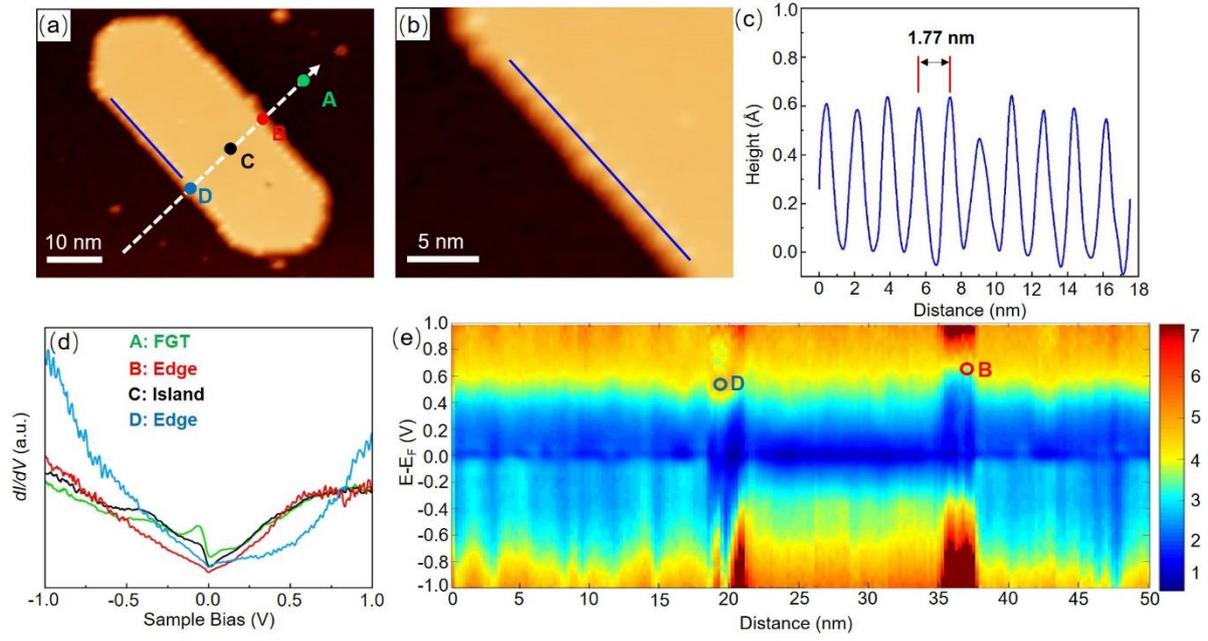

**Figure 4**. (a) Typical STM image of Bi(110) island. (b) Close-up STM image showing the reconstructions at the edge. (c) Line profile along the edge (blue solid line) in panel (b). (d) The typical *dI/dV* spectra taken at the positions indicated by the color dots, indicated in (a). (e) 2D plot of tunnelling spectra measured along the white dashed arrow line in (a).

15